\documentclass[aps,prd,twocolumn,nofootinbib,showpacs,groupedaddress]{revtex4-1}

\usepackage{amsfonts}
\usepackage{amsmath}
\usepackage{amssymb}
\usepackage{bm}
\usepackage{dcolumn}
\usepackage{graphicx}
\usepackage{graphics}
\usepackage[latin1]{inputenc}
\usepackage{latexsym}
\usepackage{rotating}
\usepackage{hyperref}
\usepackage[all]{hypcap}
\usepackage{xspace} 
\usepackage[usenames]{color}
\usepackage{mathrsfs}
\usepackage{epstopdf}

\usepackage{ulem}
\definecolor {darkgreen}{rgb}{0.2,0.7,0.2}

\newcommand{\be}{\begin{equation}}
\newcommand\ee{\end{equation}}

\newcommand\bw{\begin{widetext}}
\newcommand\ew{\end{widetext}}

\newcommand{\nn}{\nonumber}

\newcommand{\Schw}{{\mbox{\tiny Schw}}}
\newcommand{\RN}{{\mbox{\tiny RN}}}

\newcommand{\EH}{{\mbox{\tiny EH}}}
\newcommand{\Kerr}{{\mbox{\tiny Kerr}}}

\newcommand{\SR}{{\mbox{\tiny SR}}}
\newcommand{\RQG}{{\mbox{\tiny RQG}}}
\newcommand{\QG}{{\mbox{\tiny QG}}}
\newcommand{\KEF}{{\mbox{\tiny K,EF}}}
\newcommand{\KBL}{{\mbox{\tiny K,BL}}}

\begin{document}
\title{Applicability of the Newman-Janis Algorithm \\ to Black Hole Solutions of Modified Gravity Theories}

\author{Devin Hansen}
\author{Nicol\'as Yunes}
\affiliation{Department of Physics, Montana State University, Bozeman, MT 59717, USA.}

\date{\today}

\begin{abstract} 
The Newman-Janis algorithm has been widely used to construct rotating black hole solutions from non-rotating counterparts.  While this algorithm was developed within General Relativity, it has more recently been applied to non-rotating solutions in modified gravity theories.  We find that the application of the Newman-Janis algorithm to an arbitrary non-GR spherically-symmetric solution introduces pathologies in the resulting axially-symmetric metric. This then establishes that, in general, the Newman-Janis algorithm should not used to construct rotating black hole solutions outside of General Relativity.
\end{abstract}

\pacs{04.70.Bw,04.20.Jb,04.50.Kd,04.70.-s}


\maketitle
\allowdisplaybreaks
\section{Introduction}

While black holes (BHs) are theorized to be at the center of most galaxies, their direct detection remains allusive.  The evidence for their existence is indicated, for example, through the observations of S-stars around Sgr A* in the center of the Milky Way~\cite{Ghez:1998ph,Ghez:2003qj,Ghez:2008ms}, or through accretion disk observations~\cite{Narayan:1995ic,Reynolds:2002np,Shafee:2005ef,McClintock:2006xd,Brenneman:2006hw}. Direct BH detection would require observations of the BH's event horizon, perhaps through the shadows that this casts on BH accretion disks~\cite{Broderick:2005jj,Broderick:2005my,Broderick:2005xa}.  Another method to directly detect BHs is through the gravitational waves they emit when they spiral into each other and merge. After merger, the remnant rings down carrying information that would allow tests of the no-hair theorem, and thus, a verification of the BH nature of the spacetime~\cite{Dreyer:2003bv,Yunes:2013dva}. Even as a small BH spirals into a massive one, the waves emitted carry important information that can be used to map the gravitational field in the exterior of the massive BH.
	
Such observations, however, require precise knowledge of the gravitational field produced by a BH outside its event horizon. This knowledge can be obtained from solutions to the field equations, which, although easy to find analytically in General Relativity (GR), become daunting tasks in modified gravity theories. While Schwarzschild's stationary and spherically symmetric solution was derived soon after GR was first presented~\cite{Schwarzschild:1916uq}, it took about half of a century for Kerr to find an analytical axially symmetric counterpart~\cite{Kerr:1963ud}.  Similarly, stationary and spherically symmetric solutions are easy to find in modified gravity theories~\cite{yunesstein,Pani:2009wy}, but their added complexities makes axially-symmetric solutions near impossible to derive~\cite{yunespretorius,kent-CSBH,pani-quadratic}.
	
The Newman-Janis algorithm (NJA) provides a set of steps with which one can transform the  
Schwarzschild metric into the Kerr metric.  The algorithm generally proceeds as follows~\cite{NewmanJanis, DrakeSzekeres}: (i) construct a ``seed'', stationary and spherically-symmetric solution in the tetrad formalism~\cite{Newman:1961qr,Newman:1965ik,Chandrasekhar:1992bo}; (ii) complexify the tetrad; (iii) apply a complex coordinate transformation; and (iv) apply a reality condition. The original work by Newman and Janis mapped the Schwarzschild metric to the Kerr one~\cite{NewmanJanis}, but the algorithm has also been shown to work between the Reissner-Nordstr\"om metric and the Kerr-Newman metric. Since then, the NJA has found a home as a simple way to ``derive" the Kerr metric from the Schwarzschild metric. In introductory texts, the NJA is used as a means of ``rotating" the spherically symmetric, Schwarzschild metric into the complex plane to obtain the Kerr line element.

Two ambiguities are intrinsic in the NJA. The first concerns the complexification of the tetrad, i.e.~the rewriting of the real tetrad into a complex one through coordinate complexification. The second concerns the complex coordinate transformation applied to the complex tetrad. Drake and Szekeres~\cite{DrakeSzekeres} generalized this algorithm by considering a general static, spherically-symmetric seed metric, but continued to use the same complexification and coordinate transformation as in the original work of Newman and Janis~\cite{NewmanJanis}. We will here discuss these ambiguities in more detail and show that the complexification procedure is unique when considering solutions to linear order in the slow-rotation approximation. 

In recent years, the NJA has been used for two main applications: to find interior solutions that match the exterior Kerr solution~\cite{DrakeSzekeres, NJA_InteriorSolutions} and to ``rotate" spherically symmetric solutions in modified theories of gravity~\cite{Yazadjiev_NJADilationAxion, Kyriakopoulos_NJA, Dadhich:2013gr, Ghosh:2012gr, vigelandnico, Caravelli:2010cqg, johannsen-metric}. In this paper, we will be concerned with the latter. One rotates stationary, spherically-symmetric solutions of modified theories to obtain a rotating counterpart when (a) the modified theory leads to field equations that are so difficult to solve that rotating BH solutions have not been found or (b) one is considering parameterized deformations from Schwarzschild and lacks a well-defined action to prescribe these deformations. 

Until recently, case (a) was deemed unimportant because the Kerr metric was thought to be a solution of essentially all quadratic-curvature, modified gravity theories~\cite{Psaltis:2007cw}. This was so because the non-Einstein terms in the modified field equations in these theories are all proportional to the Ricci tensor or scalar, which vanishes for the Kerr metric. However, this is not the case in the more interesting dynamical Chern-Simons gravity~\cite{Grumiller:2007rv,Alexander:2009tp} and Einstein-Dilation-Gauss Bonnet theory~\cite{yunesstein}. In these theories, the field equations contain terms that are not proportional to the Ricci tensor, and thus, they are not satisfied for the Kerr metric. Finding analytic vacuum solutions that are stationary and axially symmetric in these theories is incredibly difficult, as is the case also in pure Lovelock theories in higher dimensions~\cite{Dadhich:2013gr}. This is why recent studies have considered applying the NJA to non-spinning BH solutions in pure Lovelock gravity to construct spinning ones~\cite{Dadhich:2013gr}. 

Case (b) is concerned with constructing generic spacetimes that are parametric deformations from known GR solutions, such as the Schwarzschild metric. The aim here is to carry out electromagnetic or gravitational observations in order to then place constraints on the magnitude of such deformations. Since the most general, stationary and spherically-symmetric line element has only 2 functional degrees of freedom, it is easier to parametrically deform this~\cite{Collins:2004ex,johannsen-metric} than the most general, stationary and axially-symmetric line element, which contains 4 degrees of freedom~\cite{vigelandnico}. However, since, astrophysical BHs are expected to be spinning, recent studies have considered applying the NJA to the parametrically-deformed Schwarzschild metrics~\cite{vigelandhughes,vigeland,johannsen-metric}.

In this paper, we will show that the use of the NJA for case (a) and case (b) above has no basis, and what is worse, can lead to severe pathologies in the space-time.  Hints of these pathologies can be seen in~\cite{vigelandhughes}, where an NJA rotation ``smeared" out the point singularity of a bumpy Schwarzschild metric.  Although~\cite{Pirogov:2013gr} has already argued that the NJA is ineffective at producing Kerr-like solutions in Brans-Dicke theory, this warning has fallen on deaf ears recently~\cite{johannsen-metric, Dadhich:2013gr,Yazadjiev_NJADilationAxion}. We will show explicitly that applying the NJA to non-spinning BH solutions in quadratic modified gravity, which includes Einstein-Dilation-Gauss-Bonnet theory~\cite{yunesstein}, does not lead to a spinning BH solution in this theory. We will then show that the NJA-transformed metric in fact possesses naked singularities, which renders it useless for observational studies. Although we concentrate on modified quadratic gravity, the results obtained here also apply to generic, parametric deformations of Schwarzschild line elements~\cite{Johannsen:2013rqa}. 
			
The remainder of this paper is organized as follows. Section~\ref{sec:ABC} summarizes the NJA as defined in GR. Section~\ref{sec:NJA-MG} defines the modified theory we use to test the NJA and performs the algorithm on a stationary and spherically symmetric solution in that theory. Section~\ref{sec:Analysis-Rot-Met} tests our transformed metric against an analytically-derived solution and the modified field equations for the theory used, and finds naked singularities in the spacetime. Section~\ref{sec:conclusions} concludes with a discussion of possible topics for future study.  

Henceforth, we use the conventions of~\cite{Carrol}: spacetime tensors are written with Greek indices; we use the metric signature ($-$ + + +) and geometric units with $G=c=1$. For convenience, the conversion factor between geometric units and mks units is $1M_{\odot}=1.476 \; {\rm{km}}=4.926\times 10^{-6} \; {\rm{s}}$.
	
\section{The ABC of the NJA}
\label{sec:ABC}

The NJA is a useful and simple tool to produce the rotating Kerr or Kerr-Newman metrics from the easily-derivable Schwarzschild or Reissner-Nordstr\"om line elements, respectively~\cite{NewmanJanis}.  The algorithm involves making a complex transformation on the null tetrad vectors for the non-rotating metric to arrive at the analogous rotating solution.  In this brief overview, we will largely follow ~\cite{DrakeSzekeres} and apply the algorithm to map the Reissner-Nordstr\"om to the Kerr-Newman metric. The original Schwarzschild to Kerr ``version" of the algorithm is obtained trivially by taking the limit as the BH charge vanishes.

We begin with the Reissner-Nordstr\"om metric in advanced Eddington-Finkelstein coordinates, ($v$,$r$,$\theta$,$\phi$)~\cite{DrakeSzekeres}:
\begin{equation}
ds^2_{\RN}=-fdv^2+2dvdr+r^2\left(d\theta^2+\sin^2\theta d\theta^2\right)\,,
\end{equation}
where 
\be
f \equiv \left(1-2M/r-Q^2/r^2\right)\,,
\ee
with $M$ and $Q$ the BH's mass and charge respectively. Let us rewrite the metric in the Newman-Penrose tetrad formalism~\cite{Newman:1961qr} as
\be
\label{eq:metric-in-tetrad}
g^{\mu \nu}_{\RN}=2l^{(\mu}_{\RN} n^{\nu)}_{\RN} + 2m^{(\mu}_{\RN}\bar{m}^{\nu)}_{\RN}\,,
\ee
where the overhead bar implies complex conjugate and we have defined the tetrad vectors
\begin{align}
\label{eq:tetrad-real-l}
l^\mu_{\RN}&=\left[0,1,0,0\right]\,,
\\
n^\mu_{\RN}&=\left[1,-\frac{1}{2}f,0,0\right]\,,
\\
\label{eq:tetrad-real-m}
m^\mu_{\RN}&=\left[0,0,\frac{1}{\sqrt{2}r},\frac{i}{\sqrt{2}r\sin\theta}\right]\,.
\end{align}

Let us now complexify the coordinates $(v,r,\theta,\phi) \to (\tilde{v},\tilde{r},\tilde{\theta},\tilde{\phi})$, where the overhead tilde quantities are complex. There are many ways in which this complexification can be carried out. The standard Newman-Janis choice is to replace Eqs.~\eqref{eq:tetrad-real-l}-\eqref{eq:tetrad-real-m} by
\begin{align}
\tilde{l}_{\RN}^{{\mu}}&=\left[0,1,0,0\right]\,,
\\
\tilde{n}_{\RN}^{{\mu}}&=\left[1,-\frac{1}{2}\tilde{f},0,0\right]\,,
\\
\tilde{m}_{\RN}^{{\mu}}&=\left[0,0,\frac{1}{\sqrt{2}\tilde{r}},\frac{i}{\sqrt{2}\tilde{r}\sin\tilde{\theta}}\right]\,,
\end{align}
where we have defined the complex Schwarzschild factor
\be
\tilde{f}\equiv\left[1-M\left(\frac{1}{\tilde{r}}+\frac{1}{\bar{\tilde{r}}}\right)-\frac{Q^2}{\tilde{r}\bar{\tilde{r}}}\right]\,.
\label{eqn:f-def}
\ee
Here, the overhead bar stands for complex conjugation and the Greek letters in the index lists now label complex coordinates $(\tilde{v},\tilde{r},\tilde{\theta},\tilde{\phi})$. Notice that the term proportional to $Q^{2}$ is complexified differently from the term proportional to $M$ in $\tilde{f}$. 

We now apply the complex coordinate (tilde-) transformation
\be
x^{{\rho}} \to \tilde{x}^{{\rho}} = x^{{\rho}} +ia\cos \tilde{\theta} (\delta^{{\rho}}_{\tilde{v}} - \delta^{{\rho}}_{\tilde{r}})
\label{eq: NJA}
\ee
to the complex tetrad vectors, where $a$ will later be identified with the dimensional Kerr spin parameter. As before, there is an infinite number of complex coordinate transformations that one could perform. However, only the one presented above will map the Reissner-Nordstr\"om metric to the Kerr line element. After transforming the complex tetrad vectors, we apply a reality condition that ensures that $\tilde{x}^{\rho} = \bar{\tilde{x}}^{\rho}$, simplify and obtain
\begin{align}
\tilde{l}_{\RN}^{\mu}&=[0, 1, 0, 0]\,,
\\
\tilde{n}_{\RN}^{\mu}&=\left[-1, -\frac{1}{2}\left(1-\frac{2Mr-Q}{\Sigma}\right), 0, 0\right]\,,
\\
\tilde{m}_{\RN}^{\mu}&=\frac{1}{\sqrt{2}(r+ia\cos(\theta))}\left[ia\sin(\theta), ia\sin(\theta), 1, \frac{i}{\sin(\theta)}\right]\,,
\end{align}
where
\be
\Sigma \equiv r^2 + a^2\cos^2\theta\,.
\ee

The metric resulting from the transformed null tetrad vectors will be the Kerr-Newman one in advanced Eddington-Finkelstein coordinates:
\begin{align}
g_{vv}^{\KEF}&=-f\,, \qquad g^{\KEF}_{vr}=1\,,
\\
g^{\KEF}_{v\phi}&= a\sin^2\theta \frac{2Mr-Q^2}{\Sigma}\,,
\\
g^{\KEF}_{r\phi}&= a\sin^2\theta\,, \qquad g^{\KEF}_{\theta\theta}=\Sigma\,,
\\
g^{\KEF}_{\phi\phi}&=\sin^2\theta\Phi_{\Kerr}\,,
\label{eqn:KerrAEF}
\end{align}
where
\be
\Phi_{\Kerr} \equiv \left(r^2+a^2-a^2\sin^2\theta \frac{2Mr-Q^2}{\Sigma}\right)\,.
\ee
To convert this metric to Boyer-Lindquist coordinates $(t,r,\theta,\phi)$, we perform the transformation
\begin{align}
dv &= dt + \frac{r^2+a^2}{r^2-2Mr+a^2}\,,
\label{eqn:AEF->BL}
\\
d\phi &= d\phi+\frac{a}{r^2-2Mr+a^2}dr\,,
\end{align}
resulting in the Kerr-Newman metric:
\begin{align}
\label{Kerr-BL-1}
g^{\KBL}_{tt}&=-f\,, \qquad g^{\KBL}_{t\phi}= a\sin^2(\theta)(f-1)\,,
\\
g^{\KBL}_{rr}&=\frac{\Sigma}{r^2+a^2-2Mr}\, \qquad g^{\KBL}_{\theta\theta}=\Sigma\,,
\\
g^{\KBL}_{\phi\phi}&=\sin^2\theta\Phi_{\Kerr}\,.
\label{Kerr-BL-3}
\end{align}

Although the NJA has led to the Kerr metric, the complexification and the complex coordinate transformation are somewhat unsettling. For example, the complexification $\tilde{f}$ is carried out differently for even and odd powers of radius. Similarly, the complex transformation is the identity operator plus a linear-in-spin deformation that mixes temporal-radial components and nothing else. The reason behind these choices is unclear beyond that they happen to produce the desired results. We will see that these ambiguities are at the core of the NJA failing in modified theories of gravity.

\section{The NJA in Modified Gravity}
\label{sec:NJA-MG}

The NJA is an alluring means of ``rotating'' spherically symmetric solutions into axisymmetric ones because it is simple, but does it work in modified gravity theories? While the NJA has proven to bypass much tedium when ``deriving" the Kerr solution in traditional GR, its usefulness in modified theories of gravity is often taken for granted.  This section walks through the NJA in a generic modified theory.  We show that great care must be taken when using the NJA, since generically the ``rotated" metric will contain pathologies absent in the seed metric, and thus, introduced merely by the NJA.

\subsection{BHs in Quadratic Modified Gravity}
\label{QuadG}
With gravitational wave detection imminent on the horizon, possible strong-field deviations from GR is a topic that has received much attention recently~\cite{Yunes:2013dva}.  Many modifications, including dynamical Chern-Simons gravity~\cite{CSreview}, have become popular to study such deviations.  While the familiar BH solutions of GR, ie.~the Schwarzschild and Kerr metrics, are also solution in certain theories~\cite{Psaltis:2007cw}, it has been shown in~\cite{Yunes:2011we} that these metrics are not solutions in a wide class of {\emph{quadratic gravity theories}}.

Let us consider this class of theories~\cite{Yunes:2011we}. The action adds to the Einstein-Hilbert action $S_{\EH}$ all possible quadratic combinations of curvature scalars multiplied by a scalar field $\vartheta$ and coupling constants $\alpha_{i}$:
\begin{align}
	\label{Action}
	S&\equiv S_{\EH}+\int_{\mathcal{V}} d^4x\sqrt{-g}\left\lbrace\alpha_1R^2\vartheta+\alpha_2R_{\mu\nu}R^{\mu\nu}\vartheta\right.
	\\
	&\left.+\alpha_3R_{\mu\nu\rho\sigma}R^{\mu\nu\rho\sigma}\vartheta+\alpha_4R_{\mu\nu\rho\sigma}{}^*R^{\mu\nu\rho\sigma}\vartheta\right.
	\nonumber
	\\
	&\left.-\frac{\beta}{2}[\nabla_\mu\vartheta\nabla^\mu\vartheta+2V(\vartheta)]+\mathcal{L}_{\textnormal{mat}}\right\rbrace.
	\nonumber
\end{align}
Here, $\kappa=(16\pi G)^{-1} $, $R$, $R_{\mu\nu}$ and $R_{\mu\nu\rho\sigma}$ are the Ricci scalar, Ricci tensor and the Riemann tensor respectively, $g_{\mu\nu}$ is the metric tensor and $g$ is its determinant. We can vary this action with respect to the metric and the scalar field; doing so the field equations for this theory are as follows:
\begin{align}
\label{eqn:quadgrav}
G_{\mu\nu}&+\frac{\alpha_1}{\kappa}\mathcal{H}^{(\vartheta)}_{\mu\nu}+\frac{\alpha_2}{\kappa}\mathcal{I}^{(\vartheta)}_{\mu\nu}
\\
&+\frac{\alpha_3}{\kappa}\mathcal{J}^{(\vartheta)}_{\mu\nu}+\frac{\alpha_4}{\kappa}\mathcal{K}^{(\vartheta)}_{\mu\nu}=\frac{1}{2\kappa}\left(T^{mat}_{\mu\nu}+T^{(\vartheta)}_{\mu\nu}\right),
\nonumber
\\
\label{EOM-evol}
\beta \square \vartheta &= - \alpha_{1} R^{2} - \alpha_{2} R_{\mu \nu} R^{\mu \nu} 
\nn \\
&- \alpha_{3} R_{\mu \nu \delta \sigma} R^{\mu \nu \delta \sigma} - \alpha_{4} R_{\mu \nu \delta \sigma} {}^{*}R^{\mu \nu \delta \sigma}\,,
\end{align}
where $T^{(\vartheta)}_{\mu\nu}$ is the stress energy tensor for the scalar field and ${}^{*}R_{\mu \nu \delta \sigma}$ is the dual Riemann tensor. The definitions for tensors $\mathcal{H}^{(\vartheta)}_{\mu\nu}$-$\mathcal{K}^{(\vartheta)}_{\mu\nu}$ can be found in~\cite{Yunes:2011we}.

This class of theories must be interpreted as a low-energy or curvature expansion of a more general class of theories. As such, it possesses an energy cut-off that cannot be exceeded without including higher-order operators in the action. We will here neglect such operators since the experiments and observations we have in mind are well below this cut-off. For a more detailed discussion of this class of theories as a low-energy effective model, we refer the reader to~\cite{kent-CSBH}.  

Let us now consider stationary and spherically symmetric line elements that are a small deformation from the Schwarzschild metric. For such metrics, the $\alpha_{1}$, $\alpha_{2}$ and $\alpha_{4}$ terms that source the scalar field evolution in Eq.~\eqref{EOM-evol} and the metric deformation in Eq.~\eqref{eqn:quadgrav} vanish to leading order in the deformation. The only terms that do not vanish are proportional to $\alpha_{3}$: that which sources the scalar field evolution is nothing but the Kretchmann scalar, while $\mathcal{J}^{(\vartheta)}_{\mu\nu}$ reduces to
\begin{align}
\label{eqn:j}
\mathcal{J}^{(\vartheta)}_{\mu\nu} &= 4R_{\mu \sigma \nu \delta}\nabla^{\sigma} \nabla_{\delta} \vartheta\,.
\end{align} 

With these simplifications, one can solve the modified field equations to find stationary, spherically symmetric solutions that are small deformations of the Schwarzschild metric~\cite{Yunes:2011we}. Solving the scalar field equation of motion, one finds
\begin{align}
\label{QG-field}
\vartheta^{\QG} &= \frac{\alpha_{3}}{\beta} \frac{2}{M r} \left(1 + \frac{M}{r} + \frac{4}{3} \frac{M^{2}}{r^{2}}\right) + {\cal{O}}(\zeta)\,,
\end{align}
while solving the modified field equations for the metric tensor, one finds
\begin{equation}
  ds^2_{\QG}=-f(1+h)dt^2+f^{-1}(1+k)dr^2+r^2d\Omega^2 + {\cal{O}}(\zeta^{2})\,,
  \label{eqn:NicoMetric}
\end{equation}
where\footnote{In these equations, $p$ and $q$ are the same as $\tilde{h}$ and $\tilde{k}$ in~\cite{Yunes:2011we}. We have renamed these quantities here because, for the purposes of this paper, overhead tildes are used to indicate tilde transformations.}
\begin{align}
h=&\frac{\zeta}{3f}\left(\frac{M}{r}\right)^3p\,,
\\
k=&-\left(\frac{\zeta}{f}\right)\left(\frac{M}{r}\right)^2q\,,
\\
p=&1+\frac{26M}{r}+\frac{66M^2}{5r^2}+\frac{96M^3}{5r^3}-\frac{80M^4}{r^4}\,,
\\
q=&1+\frac{M}{r}+\frac{52M^2}{3r^2}+\frac{2M^3}{r^3}+\frac{16M^4}{5r^4}-\frac{368M^5}{3r^5}\,,
\end{align}
and we have defined the dimensionless coupling constant $\zeta\equiv \alpha_3^2/(M^4\beta\kappa)$. Recall here that $f = 1 - 2 M/r$ is the Schwarzschild factor and $M$ is the ADM mass of the modified gravity BH, as discussed in~\cite{Yunes:2011we}.

\subsection{NJA on Quadratic Gravity}

We begin by transforming the metric in Eq.~\eqref{eqn:NicoMetric} to advanced Eddington-Finkelstein coordinates ($v$, $r$, $\theta$, $\phi$) via the transformation:
\begin{equation}
  dt \rightarrow dv + \left(1+\frac{2M}{r-2M}\right)dr\,,
\end{equation}
resulting in the new line element:
\begin{align}
ds^2_{\QG} &= -f(1+h)dv^2 + 2(1+h)dvdr
\nn \\  
&-\frac{(h-k)}{f}dr^2 +r^2d\Omega^2\,.
\end{align}

From here, we rewrite the metric via null tetrad vectors as in Eq.~\eqref{eq:metric-in-tetrad}. Defining $Z^\mu_a=(l^\mu,n^\mu,m^\mu,\bar{m}^\mu)$ and expanding to leading-order in $\alpha^{2}$, we find that $Z^\mu_a = Z^\mu_{a,\Schw} + \delta Z^{\mu}_{a}$, where $Z^\mu_{a,\Schw}=Z^\mu_{a,\RN}$ with $Q=0$ and
\begin{align}
\label{eqn:NPTetrad}
\delta l^\mu&=-\frac{\zeta}{2} \frac{M^{2}}{r^{2}} \frac{1}{f^{2}} \delta_{1} \delta_v^\mu
+\frac{3}{4} \zeta \frac{M^{2}}{r^{2}} \frac{1}{f} \delta_{2} \delta^\mu_r\,,
\\   
\delta n^\mu&=\frac{\zeta}{8} \frac{M^{2}}{r^2} \delta_{1} \delta_r^\mu\,,
\end{align}
with
\begin{align}
\delta_{1} &\equiv  1+\frac{4}{3}\frac{M}{r}+26\frac{M^2}{r^2}+\frac{32}{5}\frac{M^3}{r^3}+\frac{48}{5}\frac{M^4}{r^4}-\frac{448}{3}\frac{M^5}{r^5}\,,
\nn \\
\delta_{2} &\equiv 1+\frac{8}{9}\frac{M}{r}+\frac{130}{9}\frac{M^ 2}{r^2}+\frac{8}{15}\frac{M^3}{r^3}+\frac{16}{15}\frac{M^4}{r^4}-\frac{1024}{9}\frac{M^5}{r^5}\,,
\end{align}
and $\delta m^\mu  = 0$. 

The next step in the NJA demands the complexification of the coordinates, which introduces some arbitrariness. The tilde transformation $Z^\mu_a\left(x^\rho\right)\rightarrow\tilde{Z}^\mu_a\left(\tilde{x}^\rho,\bar{\tilde{x}}^\rho\right)$ used in the original NJA follows no clear repeatable form beyond that $\tilde{g}_{\mu \nu}$ must be real, and  $Z^\mu_a\left(x^\rho\right)=\tilde{Z}^\mu_a\left(\tilde{x}^\rho,\bar{\tilde{x}}^\rho\right)$ when $\tilde{x}^\rho=\bar{\tilde{x}}^\rho$.  Possible options that satisfy the aforementioned conditions include the following:
\begin{enumerate}
\item[(i)] Letting the $\alpha$-independent terms transform precisely as the NJA and the $\alpha$-dependent terms such that $r^2\rightarrow\Sigma$;
\item[(ii)] Decomposing all powers of $1/r$ into multiplicative combinations of $1/r$ and $1/r^2$ and following trends set by even and odd powers of $1/r$ in the vectors $l^\mu_\Schw$ and $n^\mu_\Schw$;
\item[(iii)] Writing all powers of $r$ in terms of $f$, and then transforming $f$ as done in the NJA.
\end{enumerate} 
All of these variations differ only by powers of $a^2/M^2$ and higher.  If one works in the slow-rotation approximation $a/M\ll 1$ and to leading order in $a/M$, these variations are irrelevant. All powers of $\mathcal{O}(a/M)$ come from the tilde-transformation of $m^\mu$ and $\bar{m}^\mu$, both of which remain unmodified in our metric. Therefore, regardless of our choice of complexification, one obtains the same rotated metric to ${\cal{O}}(a/M)$.

While any one of these methods results in a metric that in principle should be valid for all values of $a/M$, we will here expand in $a/M \ll 1$ and show only the linear-in-$a$ pieces for sake of space. After transforming back to Boyer-Lindquist type coordinates, we find that the rotated, quadratic gravity solution can be written as
\be
g_{\mu \nu}^{\RQG} = g_{\mu \nu}^{\KBL} + \delta g_{\mu \nu}^{\RQG}\,, 
\ee
where the only non-vanishing perturbations are
\begin{align}
\label{eqn:DevinMetricBL}
\delta g_{tt}^{\RQG} &= -\frac{\zeta}{3} \frac{M^{3}}{r^3} p\,,
\\
\delta g_{t\phi}^{\RQG} &= -\frac{4}{3} \zeta a \sin^2(\theta) \frac{M^{4}}{r^4} \frac{1}{f} p \,,
\\
\delta g_{rr}^{\RQG} &= - \zeta  \frac{M^{2}}{r^2} \frac{1}{f^2} q\,,
\\
\delta g_{r\phi}^{\RQG} &= -2 \zeta a \sin^2(\theta) \frac{M^{2}}{r^{2}} \frac{1}{f^{2}} q\,,
\end{align}
where recall that $g_{\mu \nu}^{\KBL}$ is the Kerr metric in Boyer-Lindquist coordinates [Eqs.~\eqref{Kerr-BL-1}-\eqref{Kerr-BL-3}], expanded in $a/M\ll 1$. 

\section{Analysis of the Rotated Metric}
\label{sec:Analysis-Rot-Met}

Now that we have constructed a rotated metric, we can begin to look more closely at its properties and determine whether or not it is a feasible metric for rotating BHs in quadratic gravity.  In order to accomplish this, the following section will be broken up into three parts.  First, we will compare this metric to the actual solution to the modified field equations to ${\cal{O}}(a/M)$ in the slow-rotation limit~\cite{pani-quadratic}.  We will then go on to evaluate the field equations for the rotated metric in the previous section to see if it is actually a solution.  Finally, we will look at some geometric and physical properties of the rotated solution.

\subsection{Comparison with a Known Solution}

Let us begin by comparing the NJA rotated metric to an analytically-derived metric for a rotating BH in modified quadratic gravity.  Pani, \textit{et.~al.}~\cite{pani-quadratic} solved the modified field equations [Eq.~\eqref{eqn:quadgrav}] in the slow-rotation approximation to $\mathcal{O}(a/M)$ to find\footnote{The conventions for the coupling constants in~\cite{pani-quadratic} differ slightly from those chosen in this paper. By comparing the action used here to that in~\cite{pani-quadratic}, we find that $\alpha^2_{3\SR}=4\alpha^2_3/(\beta\kappa)$.}
\be
\label{eqn:PaniMetric}
g_{\mu \nu}^{\SR} = g_{\mu \nu}^{\Kerr} + \delta g_{\mu \nu}^{\SR}\,,
\ee
where
\begin{align}
\delta g_{tt}^{\SR} &= -\frac{\zeta}{3}\frac{M^3}{r^3}p,
\\
\delta g_{t\phi}^{\SR} &= \frac{10}{3}\zeta a\sin^2(\theta)\frac{M^3}{r^3}\delta_\SR\,,
\\ 
\delta g_{rr}^{\SR} &= - \zeta  \frac{M^{2}}{r^2} \frac{1}{f^2} q\,,
\end{align}
and
\begin{equation}
\delta_\SR=\left(1+\frac{140M}{9r}+\frac{10M^2}{r^2}+\frac{16M^3}{r^3}-\frac{160M^4}{9r^4}\right)\,.
\end{equation}

We can now compare the rotated metric of Eq.~\eqref{eqn:DevinMetricBL} with the slowly-rotating solution of Eq.~\eqref{eqn:PaniMetric} to linear order in $a$. To $\mathcal{O}(a^0/M^0)$, both metrics reduce to that of Eq.~\eqref{eqn:NicoMetric} and obviously agree.  To $\mathcal{O}(a/M)$, however, the ($t$, $\phi$) components agree if and only if
\begin{equation}
	\frac{10}{3}\zeta a\sin^2(\theta)\frac{M^3}{r^3}\delta_\SR
	=-\frac{4}{3} \zeta a \sin^2(\theta) \frac{M^{4}}{r^4} \frac{1}{f} p \,.
\end{equation}
This, of course, is clearly not the case unless $\alpha_3=0$ or $a=0$.  Moreover, the rotated metric has a non-zero $(r,\phi)$ component, while the slowly rotating metric in Eq.~\eqref{eqn:PaniMetric} does not. Of course, this comparison assumes that both metrics are in the same coordinate system, which need not be the case. In the next section, we will show that this is not the reason for the disagreement, but rather the metrics disagree at a much more fundamental level.   

\subsection{Satisfaction the Modified Field Equations}

Even though $g_{\mu\nu}^\RQG\neq g_{\mu\nu}^\SR$ as shown in the previous subsection, these metrics still might be related by a coordinate transformation and thus $g_{\mu\nu}^\RQG$ may still be a solution to modified quadratic gravity.  Let us evaluate Eq.~\eqref{eqn:quadgrav} with the metric of Eq.~\eqref{eqn:DevinMetricBL}. First, we define 
\begin{align}
E_{\mu\nu} &\equiv G_{\mu\nu}+\frac{\alpha_3}{\kappa}\left[4R_{\mu\rho\nu\sigma}\nabla^\rho(\nabla_\sigma\vartheta)\right]
\nonumber
\\
&-\frac{1}{2\kappa}\left\lbrace T^{\textnormal{\mbox{mat}}}_{\mu\nu}+\frac{\beta}{2}\left[\nabla_\mu\vartheta\nabla_\nu\vartheta\right.\right.
\nonumber
\\
&\left.\left.-\frac{1}{2}g_{\mu\nu}\left(\nabla_\rho\vartheta\nabla^\mu\vartheta-2V(\vartheta)\right)\right]\right\rbrace\,.
\end{align}
The modified field equations are then satisfied if and only if $E_{\mu\nu}=0$ for all components of this tensor.

Let us investigate whether this is the case order by order in $a/M$. To $\mathcal{O}(a^0/M^0)$, the rotated metric reduces to the initial modified Schwarzschild metric of Eq.~\eqref{eqn:NicoMetric}, which obviously satisfies the modified field equations.  To ${\cal{O}}(a/M)$, the $(t,\phi)$ component of $E_{\mu \nu}$ is not zero, but rather
\begin{align}
E_{t\phi}&=6 \zeta a \sin^2(\theta) \frac{M^{3}}{r^5} \frac{1}{f^{2}} \left(1+24\frac{M}{r}-\frac{476}{9}\frac{M^2}{r^2}\right.
\nonumber
\\
&\left.+\frac{296}{9}\frac{M^3}{r^3}-\frac{1912}{5}\frac{M^4}{r^4}+\frac{51008}{45}\frac{M^5}{r^5}-\frac{7936}{9}\frac{M^6}{r^6}\right)\,.
\end{align}
This equation vanishes if and only if $a=0$ or $\alpha_3=0$.  

In deriving the above equation, we used the non-rotated scalar field of Eq.~\eqref{QG-field}, which obviously also matches that of~\cite{pani-quadratic} in the slow-rotation limit. Terms of ${\cal{O}}(a/M)$ in the scalar field cannot make $E_{t \phi} = 0$, as this quantity is already of ${\cal{O}}(a/M)$ and the difference between the solution found in~\cite{pani-quadratic} and the NJA rotated of Eq.~\eqref{eqn:DevinMetricBL} is also of this order. One can verify this by using an NJA rotated version of Eq.~\eqref{QG-field}. The three methods to apply the complexification of the NJA described at the end of Sec.~\ref{sec:NJA-MG} lead to the exact same scalar fields to linear order in $a/M$. With this field, one still finds that $E_{t\phi} \neq 0$. 

The NJA-rotated metric in Eq.~\eqref{eqn:DevinMetricBL} differs from the slowly-rotating solution of Eq.~\eqref{eqn:quadgrav} not because these metrics use different coordinates, but because of a fundamental issue: the NJA-rotated metric is not a solution to the vacuum modified field equations, while Eq.~\eqref{eqn:quadgrav} is. One could of course interpret the NJA-rotated metric as a {\emph{non-vacuum}} solution to the modified field equations. But as we shall show next, this ``matter'' leads to severe pathologies in the spacetime.

\subsection{Properties of the NJA-Rotated Metric}

The defining characteristic of a BH is its event horizon. The curvature singularities present in such dense objects produce a $3$-surface that bends null paths in on themselves, making observation of anything inside it impossible.  On such a surface, light cones are tilted at a $45^{\circ}$ angle, such that an observer outside the horizon can send signals to the interior but can never observe them arrive in finite proper time (and vice-versa). 

The event horizon can be defined as a null surface generated by null geodesic generators. These generators are trapped within that surface such that the normal to the surface is itself null, $n_{\mu} n^{\mu} = 0$. Let us model this surface with a scalar function $F(x^{\alpha})$, such that its normal $n^{\mu} \equiv \partial^{\mu} F$. For this surface to have a null normal, we must then have
\be
\label{eqn:EH}
g^{\mu\nu}(\partial_\mu F)(\partial_\nu F)=0\,,
\ee
We will refer to this equation as the {\emph{horizon}} equation and the radius at which $F = 0$ as the location of the event horizon. For the Kerr metric, we can take $F = r - r_{\Kerr}$, so that the horizon is located at $r = r_{\Kerr} \equiv M + (M^{2} + a^{2})^{1/2}$. 

The horizon equation can be simplified by using the symmetry properties of the spacetime. First, we note that we consider stationary and axisymmetric spacetimes, and thus, $F$ cannot depend on time or the azimuthal coordinate. With these simplifications, we can assume ring symmetry for a given latitude~\cite{Thornburg:lrr-2007-3} and Eq.~\eqref{eqn:EH} reduces to
\be
g^{rr} (\partial_rF)^2+g^{\theta\theta}(\partial_\theta F)^2=0\,.
\ee
where we have used the fact that $g^{r\theta}=0$ for the metrics considered in this paper. 

The horizon equation is thus a partial differential equation for the level surface $F$, but due to axisymmetry and reflection symmetry, $\partial_{\theta} F$ must vanish at the poles and at the equator. In this paper, it will suffice to consider the location of the horizon at the equator, $\theta = \pi/2$, for which the horizon equation reduces to the familiar
\be
\label{master-hor-eq}
g^{rr} = 0\,.
\ee
One cannot solve Eq.~\eqref{master-hor-eq} for the NJA-rotated metric, unless one expands in $a/M$. In what follows, we will {\emph{not}} use such an expansion, but rather use the full inverted metric tensor and solve Eq.~\eqref{master-hor-eq} numerically for different values of $a/M$. 

\begin{figure}[t]
\includegraphics[width=8cm,clip=true]{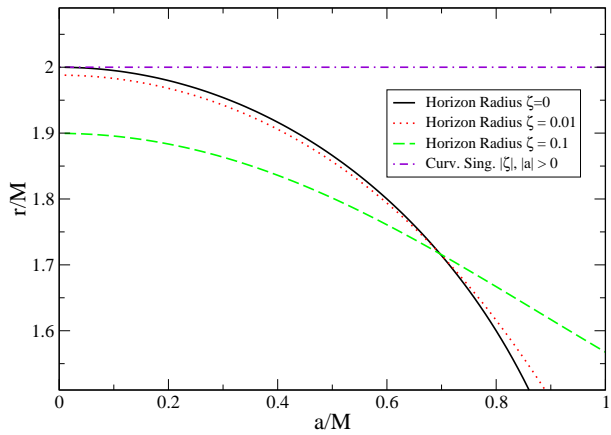}
\caption{\label{Fig:CS+EH} Properties of the NJA-Rotated metric on the equator. This figure shows the location of the event horizon at a range of $\zeta$ values (the solid line represents the GR case). Observe that the location of the curvature singularity is outside of the event horizon for all values of $a$ and $\zeta \ne 0$.}
\end{figure}

Figure~\ref{Fig:CS+EH} shows the location of the event horizon on the equator as a function of the Kerr spin parameter for different values of $\zeta$.  While the precise values were taken from version $(i)$ of the NJA, the values are nearly identical with the other versions.  Observe that larger values of $\zeta$ effectively ``dampen" the decrease in horizon radius as $a$ increases. The shift is of ${\cal{O}}(\zeta)$, however; we cannot choose values of $\zeta$ that are large, since the seed metric is a solution to the modified field equations only perturbatively in $\zeta \ll 1$. 

Let us now investigate the location of curvature singularities. A curvature singularity in the metric is a point at which the space is infinitely curved.  Such phenomenon would clash with our understanding of the observable universe if exposed, i.e.~if not hidden by an event horizon. Such exposed curvature singularities are referred to as ``naked.'' In fact, spacetimes with naked singularities are already ruled out by astrophysical observations, unless their ``infinite'' luminosity is shielded by some other process. Even from a theoretical standpoint, naked singularities are a severe pathology of any physically reasonable spacetime.

Curvature singularities can be located by finding the $3$-surfaces at which curvature scalars diverge. In this paper, we will only consider divergences of the Kretchmann scalar,
\begin{equation}
K=R_{\mu\nu\rho\sigma}R^{\mu\nu\rho\sigma}\,,
\end{equation}
which suffices to locate curvature singularities. As before, since we are considering stationary and axisymmetric spacetimes, curvature singularities are $1$-surfaces (rings), rotated along the azimuthal angle over its range and for all times. Since we wish to determine whether this curvature singularities is inside or outside of the event horizon, we further restrict attention to the equator ($\theta = \pi/2$). 

Finding the radius at which $K$ diverges for arbitrary $a/M$ is extremely difficult with the metric in Eq.~\eqref{eqn:DevinMetricBL}. This is because of how difficult it is to analytically calculate the Kretchmann scalar for this metric even when using symbolic manipulation software, such as MAPLE with the GRTENSORII package~\cite{grtensor}. We will thus restrict attention to expansions in $a/M \ll 1$ (see Appendix~\ref{Small-a-exp}). In particular, we will calculate the full Kretchmann scalar to ${\cal{O}}(a^{4}/M^{4})$ and ${\cal{O}}(a^{6}/M^{6})$ and find solutions to $1/K=0$ also to these orders, using the latter as an estimate of the error in the former. 

\begin{figure}[t]
\includegraphics[width=8.5cm,clip=true]{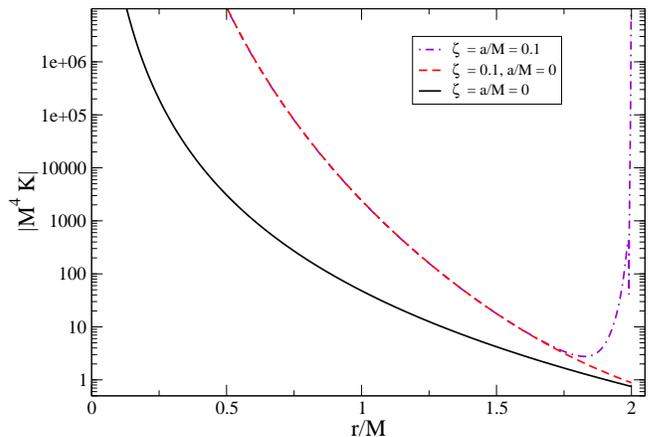}
\caption{\label{Fig:KDiv} Kretchmann scalar as a function of radius on the equator for different values of $\zeta$ and $a/M$. Observe that while all curves are singular at $r/M=0$, when $a$ and $\zeta$ are simultaneously nonzero, the Kretchmann also diverges at $r/M=2$. Observe also that when $\zeta \neq 0$, but $a/M=0$, the Kretchmann still only diverges at $r/M = 0$.}
\end{figure}

We find that the NJA rotated metric possesses two curvature singularities when $\zeta \neq 0$ \emph{and} $a \neq 0$ at the equator: one of them is always at $r/M = 0$. The second occurs only when $a \neq 0$ and $\zeta \neq 0$, and is located at $r/M=2$. Figure~\ref{Fig:KDiv} shows the Kretchmann as a function of radius for different choices of $a/M$ and $\zeta$. Clearly, when both $a$ and $\zeta$ are simultaneously nonzero, an additional singularity in the Kretchmann scalar appears at $r/M = 2$. We have verified that the Kretchamnn indeed diverges at these radii through an analytical calculation that we present in Appendix~\ref{Small-a-exp}.  

While the $r/M = 0$ singularity is not a problem, given that it is inside the event horizon, the $r/M = 2$ curvature singularity is. If one looks at the $\zeta=0.1$ curve in Fig.~\ref{Fig:CS+EH}, one finds that $r/M = 2$ is clearly above the event horizon curve for all values of $0<a/M \ll 1$. In the limit $a \to 0$, this singularity disappears. We show see this explicitly in Appendix~\ref{Small-a-exp}. This result is consistent with that of~\cite{yunesstein}, who already showed that the $\zeta \ll 1$ expanded, spherically symmetric modified metric does not possess curvature singularities outside the event horizon. We see then that naked singularities are a generic property of the NJA rotated metric in the slow-rotation limit. 

One may wonder whether these curvature singularities are also present in the slowly-rotating metric of \cite{Pani:2009wy}, or whether they are actually introduced by the Newman-Janis algorithm. To figure this out, one can compute the Kretchmann scalar for the metric of~\cite{Pani:2009wy} and find the radii at which its inverse vanishes. We have done this calculation and found that this Kretchmann scalar is identically equal to the one computed with the spherically symmetric metric of~\cite{yunesstein}, when expanded to ${\cal{O}}(a/M)$. Such an expansion is crucial since the metric of~\cite{Pani:2009wy} is only valid to linear order in $a/M$. Therefore, the metric of~\cite{Pani:2009wy} contains curvature singularities at exactly the same locations as the spherically symmetric metric of~\cite{yunesstein}. It then follows that the curvature singularities showed in Fig.~\ref{Fig:CS+EH} have nothing to do with slowly-rotating BHs in modified quadratic gravity, but rather are artificially introduced by the Newman-Janis algorithm.

\section{Implications and Discussion}
\label{sec:conclusions}

We have shown that, in general, the NJA is not appropriate to generate rotating BH solutions in modified gravity theories. We have considered a class of such theories, quadratic gravity, in which both an exact stationary and spherically symmetric BH solution is known, as well as a stationary and axisymmetric BH solution in the slow-rotation approximation. We then applied the NJA to the former and found that it differs from the latter. In fact, the rotated NJA metric was shown to not be a vacuum solution of this theory at all. Finally, the ``matter'' introduced by the NJA rotation was shown to introduce severe pathologies in the spacetime, ie.~naked singularities. 

We hope this paper acts as a word of caution when using the NJA to generate rotating BH metrics. In recent years, the NJA has been used to find rotating BHs in pure Lovelock theories~\cite{Dadhich:2013gr}, as well as to create parametrically-deformed, rotating BHs~\cite{johannsen-metric}. As in the Brans-Dicke case~\cite{Pirogov:2013gr}, we have shown here that the NJA does not necessarily preserve the field equations. What is worse, the NJA tends to introduce pathologies in the spacetime, as found also recently in~\cite{Johannsen:2013rqa} when considering parametric Schwarzschild deformations. 

In particular, we find that the NJA generically introduces naked singularities, which immediately implies that NJA-rotated parametrically deformed spacetimes are {\emph{not}} suitable for observational studies (be it electromagnetic or gravitational). When such metrics are used for accretion disk studies, the trajectory of gas across the naked singularity would lead to infinite luminosities and a potential breakage of causality. Similar problems would be faced if such metrics were used to evolve extreme mass-ratio inspirals, when considering gravitational wave observations.

Our study, however, does not rule out the existence of a modified NJA that when applied to the spherically-symmetric solution of a given set of modified field equations would yield its axisymmetric counterpart. The most likely piece of the algorithm to modify is the complex coordinate transformation of the tetrad. This modification, however, would probably have to be derived on a per theory basis, and thus, would not be generic. Such a derivation would require {\emph{a priori}} knowledge of the rotating BH solution in the particular modified theory under consideration. But if such a solution is known, the usefulness of the modified NJA would be minimal.  

\section{Acknowledgments}
We would like to thank Scott Hughes for reviewing our manuscript and providing us with useful feedback and insights. DH thanks the Undergraduate Gravity Research Program of the MSU gravity group, under which this study was completed. NY acknowledges support from NSF grant PHY-1114374, as well as support provided by the National Aeronautics and Space Administration from grant NNX11AI49G, under sub-award 00001944, and also the NSF CAREER Award PHY-125063. Some calculations used the computer-algebra system MAPLE, in combination with the GRTENSORII package~\cite{grtensor}.

\appendix
\section{Slow-Rotation Expansion}
\label{Small-a-exp}

In this appendix, we provide analytic expressions for the Kretchmann scalar, the radius at which this quantity diverges and the radius at which the horizon equation [Eq.~\eqref{master-hor-eq}] is satisfied. Because of the analytical complexity of these quantities, we expand them in $a/M \ll 1$ and work to fourth order in this small quantity. Moreover, we also provide expressions only along the equatorial plane, ie.~$\theta = \pi/2$.   

The Kretchmann scalar to ${\cal{O}}(a^{4}/M^{4})$ at $\theta = \pi/2$ is given by 
\begin{widetext}
\begin{align}
	K&=48\frac{1}{\bar{r}^6}+\frac{\zeta}{5\bar{r}^{12}}\left(-160\bar{r}^5-80\bar{r}^4-11520\bar{r}^3-1120\bar{r}^2-2048\bar{r}+114400\right)
	\nonumber
	\\
	&-\frac{2}{15}\frac{\zeta\bar{a} ^2}{\bar{r}^{13}(\bar{r}-2)^3}\left(870\bar{r}^7+515\bar{r}^6+56890\bar{r}^5-127585\bar{r}^4+71908\bar{r}^3-761136\bar{r}^2+214876\bar{r}-1563520\right)
	\nonumber
	\\
	&-\frac{2}{15}\frac{\zeta\bar{a}^4}{\bar{r}^{14}(\bar{r}-2)^4}\left(1260\bar{r}^7+1365\bar{r}^6+73970\bar{r}^5-157422\bar{r}^4+93676\bar{r}^3-1173136\bar{r}^2+3326688\bar{r}-2468480\right)
	\nonumber \\
	&+ {\cal{O}}(\zeta^{2},\bar{a}^{6})\,,
	\label{eq:kretchmann}
\end{align}
where we have defined $\bar{r} \equiv r/M$ and $\bar{a} \equiv a/M$. One clearly sees that Eq.~\eqref{eq:kretchmann} diverges at $\bar{r} = 2$ only when $\zeta \neq 0$ and $\bar{a} \neq 0$. Moreover, one also sees that when $\zeta = 0$ and $\bar{a} = 0$, then the only divergence occurs at $\bar{r} = 0$. In general, solving the equation $1/K=0$ in an expansion about $a/M \ll 1$, we find the two solutions
\begin{align}
\bar{r}_{\rm{sing}} & = {\cal{O}}(\zeta^{2},\bar{a}^{6}) \,,
\qquad {\rm{for}} \; {\rm{all}} \quad \bar{a}\,,
\nonumber \\
\bar{r}_{\rm{sing}}  &= 2 + {\cal{O}}(\zeta^{2},\bar{a}^{6})\,,
\qquad {\rm{provided}} \quad \bar{a} \neq 0 \quad {\rm{and}} \quad \zeta \neq 0\,.
\end{align}
Keeping the expansion of $K$ to higher-order in $\bar{a}$ does not change the above results. We have checked this by computing the Kretchmann scalar to ${\cal{O}}(\bar{a}^{6})$. This is because the $\bar{a}^{6}$ term left out of Eq.~\eqref{eq:kretchmann} can never cure the divergence of the terms proportional to $\bar{a}^{2}$ at $\bar{r} = 2$, unless $\bar{a}$ is not much smaller than unity. 

The horizon equation is simply
\begin{align}
		1&-\frac{2}{\bar{r}}+\frac{\bar{a}^2}{\bar{r}^2}
		+\frac{12}{5} \frac{\zeta}{ \bar{r}^7}\left[\left(-\frac{140}{9}+\bar{r}+\frac{2}{3}\bar{r}^2+\frac{5}{36}\bar{r}^4+\frac{65}{24}\bar{r}^3+\frac{5}{48}\bar{r}^5\right)\right.
		\nonumber
		\\
		&\left.+\frac{4}{3}\left(-\frac{320}{3}+\bar{r}+\frac{1}{2}\bar{r}^2+\frac{5}{6}\bar{r}^4+\frac{325}{24}\bar{r}^3+\frac{15}{16}\bar{r}^5\right)\right]= {\cal{O}}(\zeta^{2})\,.
	\end{align}
This equation is valid to all orders in $\bar{a}$. We have solved this equation numerically for $\bar{r}$, given different values of $\bar{a}$ and $\zeta$, which we plotted in Fig.~\ref{Fig:CS+EH}. 
\end{widetext}

\bibliography{master}
\end{document}